\AddToHook{package/hyperxmp/before}{\RequirePackage[bookmarksnumbered,unicode]{hyperref}}
\documentclass[sigconf]{acmart}
\usepackage{booktabs,tabularx,microtype,balance}

\definecolor{ink}{HTML}{303B45}
\definecolor{nativeblue}{HTML}{156B91}
\definecolor{codeletteal}{HTML}{187D73}
\definecolor{radioorange}{HTML}{B66B13}
\definecolor{noisepurple}{HTML}{7A4E9E}
\newcolumntype{Y}{>{\raggedright\arraybackslash}X}
\newcommand{\yes}{\textbullet}
\hypersetup{pdftitle={Controlled Radio Fault Injection and Telemetry Evaluation for Open 5G RANs},pdfsubject={ZMQ radio fault injection and 5G telemetry evaluation}}

\copyrightyear{2026}
\acmYear{2026}
\setcopyright{cc}
\setcctype{by-nc-nd}
\acmConference[MobiHoc '26]{The Twenty-seventh International Symposium on Theory, Algorithmic Foundations, and Protocol Design for Mobile Networks and Mobile Computing}{November 23--26, 2026}{Tokyo, Japan}
\acmBooktitle{The Twenty-seventh International Symposium on Theory, Algorithmic Foundations, and Protocol Design for Mobile Networks and Mobile Computing (MobiHoc '26), November 23--26, 2026, Tokyo, Japan}
\acmDOI{10.1145/3842721.3856774}
\acmISBN{979-8-4007-2975-1/2026/11}

\input{arxiv_notice_mobihoc.tex}

\begin{document}
\title[Controlled Radio Fault Injection and Telemetry Evaluation]{Controlled Radio Fault Injection and Telemetry Evaluation for Open 5G RANs}
\author{Babak Azkaei}
\email{b.azkaei@tue.nl}
\affiliation{\institution{Eindhoven University of Technology}\city{Eindhoven}\country{The Netherlands}}
\author{Maxim B\^{i}tiu}
\email{m.bitiu@student.tue.nl}
\affiliation{\institution{Eindhoven University of Technology}\city{Eindhoven}\country{The Netherlands}}
\author{Georgios Exarchakos}
\email{g.exarchakos@tue.nl}
\affiliation{\institution{Eindhoven University of Technology}\city{Eindhoven}\country{The Netherlands}}
\author{Kishor Chandra Joshi}
\email{k.c.joshi@tue.nl}
\affiliation{\institution{Eindhoven University of Technology}\city{Eindhoven}\country{The Netherlands}}

\renewcommand{\shortauthors}{Azkaei et al.}

\begin{abstract}
Radio access network (RAN) telemetry reveals radio conditions, protocol behavior, and their consequences for applications. Evaluating its usefulness requires controlled impairments and observations of the same execution. We present C and Python channel brokers and a fault injection workflow for software 5G testbeds connecting an OCUDU base station and a software UE through ZeroMQ. The brokers apply scheduled, directional transformations to I/Q samples and restore the link, while the workflow jointly records applied faults, native and programmable (codelet) telemetry, and application delay.
We demonstrate this capability through 45 comparative trials of brief uplink interruptions and stepped additive noise. These use cases distinguish recognizing a radio change from measuring its service consequence: native missing-feedback events retrospectively identify interruptions, whereas noise changes decoding-error measurements without producing those events. Median on-time round-trip times remain near 30~ms, but the prespecified comparisons show mean increases in on-time 95th-percentile delay of 354 and 604~ms for one and five interruptions, and 113~ms for noise. Adding receive-side radio link control (RLC) codelet summaries did not improve interruption classification on held-out trials. The framework provides a reusable way to evaluate measurement choices under known radio faults, before using them in monitoring or learning systems.
\end{abstract}

\maketitle

\begin{figure*}[t]
\centering
\includegraphics{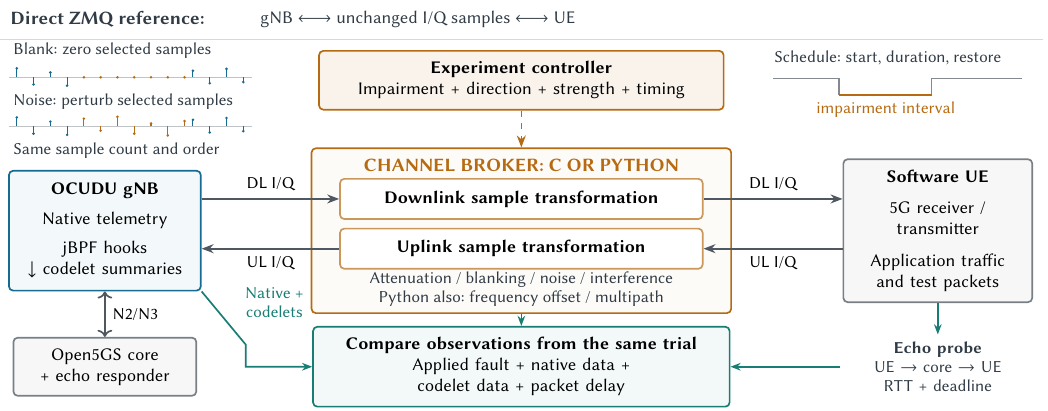}
\caption{From a direct-ZMQ link to controlled radio experiments. The controller schedules a sample transformation in either direction, and the selected C or Python broker applies it before reception. Applied-fault records, native telemetry, codelets, and a UE--core--UE echo probe describe the same run. The two brokers are alternatives, not serial stages.}
\Description{An experiment controller supplies impairment type, direction, strength and timing to one broker between OCUDU and the software UE. Separate uplink and downlink paths preserve sample transport. Native and codelet outputs, broker records, and echo-probe outcomes feed joint analysis. Blank and noise examples show zeroed or perturbed samples of unchanged count and order.}
\label{fig:architecture}
\end{figure*}

\section{Introduction}
A radio problem raises two monitoring questions: what changed in the link, and what did that change mean for applications? These questions matter for AI-assisted radio access networks (AI-RAN), whose learning and control functions depend on telemetry. The IMT-2030 vision places artificial intelligence alongside communication~\cite{imt2030}, but collecting more measurements does not by itself establish that they capture the behavior a monitoring task needs.
Open RAN provides several observation paths. Native base-station reports and event logs expose implementation-defined measurements, while standardized interfaces offer defined measurement services~\cite{native,oran}. Programmable telemetry adds small verified programs, or \emph{codelets}, that compute user-defined summaries at hooks inside a running RAN function~\cite{janus}. Comparing these alternatives requires more than a common data collector: researchers also need a known change to the radio conditions and an independent observation of its consequences.

Our starting point is a software 5G testbed in which an OCUDU base station (gNB) and software user equipment (UE) exchange digital in-phase/quadrature (I/Q) samples over ZeroMQ (ZMQ)~\cite{zmq4g,srsue}. The selected direct-link configuration has no intervening channel process. We add a \emph{channel broker} that transforms samples before reception, together with a workflow that schedules the change, restores the link, and collects observations from the same run (Figure~\ref{fig:architecture}). This permits controlled experiments without radiating interference. It complements over-the-air testing, where real propagation is exercised but surrounding activity and movement may be harder to repeat.

The main contribution is this reusable experimental capability. The telemetry study demonstrates its use. We contribute:
\begin{enumerate}
\item C and Python channel brokers for directional, sample-scheduled impairments while preserving ZMQ sample transport (\S\ref{sec:brokers}).
\item A fault injection workflow that connects applied and restored changes to native, codelet, and application observations (\S\ref{sec:workflow}--\ref{sec:measure}).
\item Interruption and additive-noise use cases showing why fault identification, service delay, and internal protocol timing need different measurements (\S\ref{sec:design}--\ref{sec:results}).
\end{enumerate}

We have released the brokers and fault-control tools as \emph{OCUDU-ZMQ-Radio-Fault}\footnote{\url{https://github.com/BabakAz/OCUDU-ZMQ-Radio-Fault}}.

\section{Radio-Fault Injection Framework}
\label{sec:framework}
\subsection{Brokers at the Virtual-Radio Boundary}
\label{sec:brokers}
The testbed comprises the OCUDU gNB, an Open5GS core, and srsUE. A ZMQ request--reply connection in each direction carries I/Q samples between gNB and UE. Either broker interposes at this boundary, transforms selected samples, and forwards a block of the original length. The existing receiver, scheduler, link adaptation, and retransmission logic then respond to the modified signal. Unlike IP-layer delay or loss injection, the broker does not directly delay or discard application packets.

The C broker is a compact relay for attenuation, blanking, complex Gaussian noise, and a sinusoidal interferer. The Python broker uses a headless NumPy processing path with optional GNU Radio visualization; it also supports carrier-frequency offset and selected 3GPP tapped-delay-line (TDL) channels~\cite{tdl}. Both expose the same experimental controls: impairment, direction, strength, start, and restoration. Table~\ref{tab:mechanisms} separates their supported operations from the evidence reported here. The repeated studies use C; the Python backend broadens the available signal transformations rather than serving as a performance baseline.

\begin{table}[t]
\caption{Broker capabilities and reported evaluation. R: repeated comparison in this paper, using C; F: separate full-stack functional checks; --: no full-stack evaluation reported. Implementation support does not imply equivalent validation of both backends.}
\label{tab:mechanisms}
\centering\small
\begin{tabularx}{\columnwidth}{@{}lYccc@{}}
\toprule
Mechanism & Sample operation & C & Python & Study\\
\midrule
Attenuation & Amplitude scaling & \yes & \yes & --\\
Blank & Zero selected samples & \yes & \yes & R\\
Noise & Complex Gaussian addition & \yes & \yes & R\\
Tone & Sinusoidal addition & \yes & \yes & F\\
Freq.\ offset & Phase rotation & & \yes & F\\
Multipath & Weighted delayed paths & & \yes & F\\
\bottomrule
\end{tabularx}
\end{table}

These operations are controlled digital stressors, not complete replicas of physical faults. A tone is not an interfering cell, blanking is not a complete blockage model, and the noise setting is referenced to digital power rather than a calibrated receiver signal-to-interference-plus-noise ratio (SINR). The static TDL examples omit mobility and spatial effects. Direction is nevertheless important: the uplink carries both UE data and control, including buffer-status reports (BSRs) and feedback for downlink transmissions~\cite{mac38321}. An uplink-only change can therefore affect downlink feedback even when downlink samples are unchanged.

\subsection{Scheduling a Change and Observing Restoration}
\label{sec:workflow}
An impairment schedule specifies when each transformation starts and ends, counted in samples. For example, at 23.04~Msamples/s a 10-ms blank replaces 230,400 samples with zeros. When a boundary falls inside a ZMQ message, the broker applies the relevant settings to each part and still returns one message of unchanged length. The receiver encounters a missing signal, not a relay that has stopped replying. Noise-generator and tone-phase states persist across changes, so repeatability also depends on input history and initial state.

The workflow proceeds from preparation to application and then observation. It waits for UE attachment and capture readiness, activates the schedule, and records when the broker processes, forwards, and restores each transformation. These records distinguish the requested change from its application. A broker-process test suite checks sample preservation, boundaries, restoration, and shutdown in 34 cases across the two backends; separate full-stack checks exercise the operations marked F in Table~\ref{tab:mechanisms}.

Restoration is deliberately limited to the signal transformation. It does not restart the gNB or UE, clear queues, or reset link adaptation. Their response may therefore outlast the injected fault. The workflow continues observing after restoration, allowing the researcher to distinguish \emph{how long the signal was changed} from \emph{how long application behavior remained different}. Replaying a schedule controls the intervention, not an identical subsequent protocol trajectory.

\subsection{Three Complementary Measurement Paths}
\label{sec:measure}
The workflow records three views of the same execution. \emph{Native telemetry} comprises the gNB's periodic reports, scheduler events, and physical-layer logs. \emph{Codelets} compute internal summaries through Janus/jBPF; we adapted and extended Microsoft's \texttt{jrtc-apps} hooks and examples for OCUDU~\cite{janus,jrtcapps}. An independent \emph{UDP echo probe} sends requests from the UE to a responder beside the core and receives them back, measuring round-trip time (RTT) and deadline misses. Applied-fault records identify the intervention, while the probe supplies a service outcome. Neither is an input to the telemetry classifiers.

These paths are complementary rather than interchangeable. Native feedback can reveal a disruption; a codelet can summarize a particular internal event; the echo probe measures the complete request--reply path. Researchers can hold the fault schedule fixed while changing the observation point or summary they evaluate. The use cases below apply this workflow to radio-event identification, application delay, and receive-side radio link control (RLC) timing.

\begin{figure*}[t]
\centering
\includegraphics{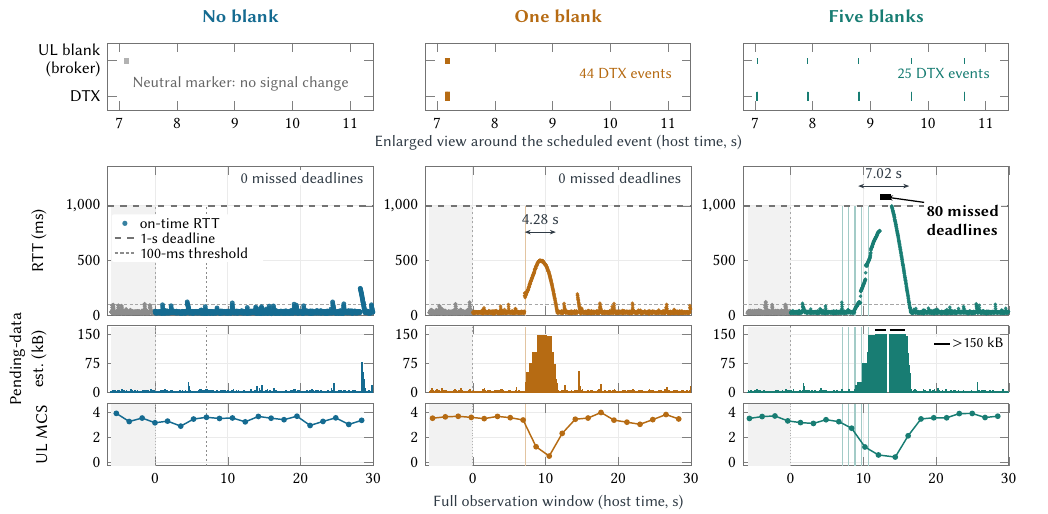}
\caption{A complete matched interruption block, selected after collection for illustration. Top: broker blank spans and native DTX events on the enlarged host-time axis. Lower rows show on-time RTT at request send time, pending-data estimates (maximum per 100~ms), and report-mean uplink MCS. The black bar marks 80 deadline-missed requests, not RTT; arrows span consecutive slow requests, not recovery time. Pending estimates above the 150-kB display limit use a dark line.}
\Description{Three trial columns compare no interruption, one blank, and five blanks. Native feedback forms zero, one and five groups. The one-blank and five-blank trials have 4.28-second and 7.02-second slow-reply runs; only the five-blank trial has 80 deadline misses. The pending-data and MCS traces show a longer high-demand, low-MCS interval with five blanks. Quantized scheduler estimates are not physical queue measurements.}
\label{fig:incident}
\end{figure*}

\section{Evaluation Design}
\label{sec:design}
\subsection{Two Radio-Fault Use Cases}
We use two deliberately different interventions: brief interruptions that remove the desired signal, and stepped additive noise that leaves it present. They test whether measurements that identify one fault also describe a different disturbance and its service consequence.

A single Linux VM hosts OCUDU~26.04 with our instrumentation and compatibility patches, Open5GS~2.8.0, srsUE, the C broker, and collectors. One UE uses a 20-MHz band-3 cell, 15-kHz subcarrier spacing, 64-QAM modulation-and-coding-scheme (MCS) tables, and 23.04~Msamples/s. Background UDP traffic is fixed at 2~Mbit/s per direction. The observed uplink SINR is about 1~dB without injection; relaxed radio-link-failure limits keep the connection active, so failure-declaration timing is not evaluated.

\textbf{Interruptions.} Each matched block runs, in random order, \emph{No blank}, \emph{One blank} of 50~ms, and \emph{Five blanks} of 10~ms with starts 0.5~sample-seconds apart. The first blank starts four sample-seconds after activation. The fault conditions suppress the same total number of samples, while the downlink stays unchanged. The control executes a neutral marker at the first blank's position.

\textbf{Stepped noise.} Separate randomized pairs compare \emph{Noise control} with an uplink additive white Gaussian noise (AWGN) staircase. Starting four sample-seconds after activation, three 0.5-sample-second stages add noise of power $0.01P_{\mathrm{ref}}$, $0.1P_{\mathrm{ref}}$, and $P_{\mathrm{ref}}$, then restore noise-free forwarding. Desired-signal gain remains one. The fixed reference $P_{\mathrm{ref}}$ is the mean input energy per sample of a qualification prefix including startup and attachment.

\subsection{Observations and Comparisons}
\label{sec:analysis}
\begin{figure*}[t]
\centering
\includegraphics{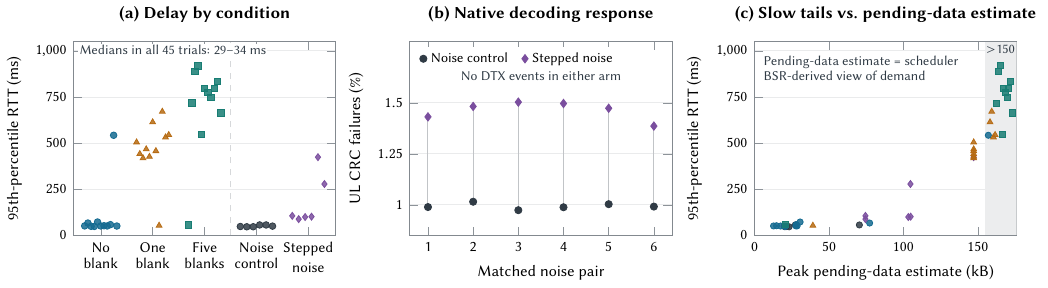}
\caption{Three views of the 45 trials. (a) Each marker is one trial's on-time p95 RTT; its small horizontal offset retains block or pair order. (b) Noise increases native decoding-error fractions without DTX events. (c) Larger pending-data estimates accompany higher p95 RTTs in a post-hoc comparison. Values above the open-ended 150-kB BSR level are spread within the shaded band for visibility, not plotted as exact queue sizes.}
\Description{Three panels show all individual trial p95 RTTs, paired noise-control and noise CRC-failure fractions, and the association between pending-data estimates and RTT p95. Different markers identify conditions. All observations remain present, including small fault responses and a high-delay control. Above-150-kilobyte pending estimates form an open-ended category rather than exact queue measurements.}
\label{fig:distributions}
\end{figure*}

We prespecified ten interruption blocks and five noise pairs. After inspecting their results, we added one block and one pair. All 45 trials appear in Figure~\ref{fig:distributions}; Table~\ref{tab:service} uses only the prespecified comparisons. Each trial starts fresh gNB, UE, broker, probe, and capture processes; the core is shared within a batch. We record at least five host seconds of baseline and analyze the 30 host seconds after the activation acknowledgment.

The probe sends 128-byte requests at 50~Hz with a one-second deadline. We report each trial's median and 95th-percentile (p95) RTT among replies received on time and before the observation cutoff. The p95 describes the slow end of those successful replies, not the delay of unanswered requests. Deadline misses, skipped send opportunities, and unresolved requests remain separate. For service effects, we subtract each control's p95 from its matched fault trial's p95, then average the differences. The 95\% percentile intervals use 20,000 bootstrap resamples of whole blocks or pairs, not individual packets.

Fault schedules use sample time; RTT and deadlines use host elapsed time. ZMQ sample progression need not match wall-clock speed~\cite{srsue}, and we do not convert RTT using an assumed rate. Broker processing, native-event arrival, and probe send/receive records retain their own timestamp meanings. The detailed incident is a post-hoc illustration, not a model of packet-level recovery. Its marked slow-request runs use consecutive requests above 100~ms RTT or missing a deadline within 12 host seconds of the first blank or neutral marker.

\subsection{Native and Programmable Telemetry}
Native discontinuous-transmission (DTX) events here indicate missing usable uplink feedback for downlink hybrid automatic repeat request (HARQ) transmissions. A pilot-informed rule, fixed before the campaign, groups events separated by at most 100~ms at the collector. Zero, one, or multiple groups classify completed interruption trials; this is not online detection or identification of a physical fault cause. We also examine native cyclic redundancy check (CRC) failures, MCS, and the scheduler's BSR-derived estimate of pending uplink data.

The selected RLC codelets report the count, mean latency, and maximum
latency of accepted completed uplink service data units (SDUs). Timing starts at reception of an SDU or its first retained segment and ends at completed reassembly, excluding earlier UE waiting and the return path. The standardized measurement \texttt{DRB.RlcDelayUl}, not captured here, already exposes a mean~\cite{ts28552}; the codelets add per-bearer extrema and collection context relative to our captured native outputs. Their intervals are not assigned exact host-time fault overlap without a verified producer-clock mapping.

\textbf{Experiment accounting.} Native and service data qualify in all 45 trials. Four RLC vectors fail the collection checks, leaving 41 complete vectors. The frozen native-versus-native-plus-RLC comparison trains depth-two trees on the first five interruption blocks and tests on the next five, with identical eligible rows, training-only preprocessing, 14 shared training cases, and 13 shared test cases. Missing cases are not replaced. Detailed qualification records remain separate from the comparisons.

\section{Results: Two Uses of the Framework}
\label{sec:results}
\subsection{Interruptions: From Radio Events to Service Delay}
The native rule identifies the programmed pattern in all 33 interruption trials: no DTX groups without a blank, one with a single blank, and five with separated blanks. Thresholds from 20 to 500~ms give the same grouping on the 30 prespecified interruption trials. Native feedback therefore resolves these particular patterns retrospectively. The event count is not a severity measure: one blank produces more DTX events than five, despite a smaller average effect on the slow-reply tail.

Figure~\ref{fig:incident} makes the distinction visible in one complete matched block. The top row locates the applied blanks and missing feedback; the lower rows follow application delay, pending data, and uplink MCS. The one-blank trial contains a 4.28-second span of consecutive slow requests. With five blanks, the corresponding span is 7.02~seconds and includes 80 missed deadlines. These are spans of request send times, not measured recovery times or the RTT of one packet. In both fault trials, reduced MCS and elevated pending-data estimates accompany the delayed replies. The joint record supports examining these relationships without assigning each packet's waiting time to a particular mechanism.

\subsection{Repeated Service Outcomes}
\label{sec:service}
Across all 45 trials, median on-time RTT remains near 30~ms, but p95 varies substantially (Figure~\ref{fig:distributions}a). Table~\ref{tab:service} quantifies the matched effects in the prespecified comparisons: one and five blanks increase p95 by 354 and 604~ms on average. The pattern concerns slow replies rather than a uniform delay increase for every request.

\begin{table}[t]
\caption{Mean increase in on-time p95 RTT over the matched control. Estimates use the prespecified comparisons only: ten interruption blocks and five noise pairs. Intervals resample whole matched units; Figure~\ref{fig:distributions} separately shows all 45 trials.}
\label{tab:service}
\centering\small
\begin{tabular}{@{}lrrr@{}}
\toprule
Fault & Matched units & Increase (ms) & 95\% interval (ms)\\
\midrule
One blank & 10 blocks & 354 & [225, 468]\\
Five blanks & 10 blocks & 604 & [432, 746]\\
Stepped noise & 5 pairs & 113 & [45, 241]\\
\bottomrule
\end{tabular}
\end{table}

The two blanked trials with low p95 RTTs also show smaller pending-data estimates and milder MCS changes. The high-delay control shows gaps in probe sending. On-time RTT must be read alongside service availability: across the full dataset, no-blank and five-blank trials contain 96 and 545 missed deadlines, respectively, while the other conditions have none. The probe skips 149 opportunities and leaves 59 sends unresolved at the cutoff. These outcomes are not assigned an artificial RTT or silently removed from the record; complete per-trial accounting is retained with the data.

\subsection{Noise: A Different Radio Response}
\label{sec:noiseresults}
The stepped-noise experiment demonstrates why the framework supports more than interruption injection. Neither arm of any noise pair produces DTX events. The interruption rule alone therefore cannot distinguish them, although other native measurements respond: the uplink CRC-failure fraction rises in every pair (Figure~\ref{fig:distributions}b). Application p95 also rises in each pair, by 42--367~ms, with a mean increase of 113~ms in the five prespecified pairs. Median RTT remains around 30~ms and neither arm misses a reply deadline.

This is not evidence that native telemetry misses noise. Rather, different native observations describe different responses: DTX exposes absent feedback, while CRC results describe decoding outcomes. The noise and interruption studies also have different durations and signal transformations, so their delay effects do not share a calibrated severity scale. Reusing the broker interface and observation workflow makes such measurement comparisons possible without redesigning the experiment for each fault.

\subsection{Choosing Measurements for the Question}
In the exploratory, all-trial view of Figure~\ref{fig:distributions}c, larger pending-data estimates generally accompany higher p95 RTTs. This adds context beyond identifying the injected condition. The estimate is the scheduler's quantized, BSR-derived view of demand; the highest short-BSR category is open-ended above 150~kB~\cite{mac38321}. It is not a physical queue measurement or an individual packet's waiting time, and the association is not a validated alarm threshold or proof of where the delay occurred.

The RLC codelets observe a different part of delivery. Their means remain approximately 4.5--4.9~ms, while their maxima vary across conditions. Receive-side timing starts only after an SDU or its first retained segment reaches the gNB; it cannot include all earlier waiting or the rest of the echo path. On the frozen shared holdout, native-only and native-plus-RLC trees correctly classify all 13 eligible shared test cases, yielding no accuracy increase on this task. Both trees use a native HARQ-failure fraction. This bounded result does not establish general codelet redundancy: it shows that an additional internal summary need not improve a distinction already resolved by the native inputs.

\section{Practical Value and Scope}
\label{sec:lessons}
The framework turns telemetry selection into a controlled experiment. A researcher can choose a signal change, record its application and restoration, and ask whether a measurement reveals the radio response, the application consequence, or a particular internal stage. The interruption and noise studies demonstrate these distinct questions: a feedback pattern identifies one intervention, decoding errors respond to another, and neither alone describes every aspect of service delay.

Programmability is useful when it addresses a specific missing observation. For example, investigating uplink waiting would motivate UE enqueue/transmit timestamps and time-local grant or demand summaries, whereas a reassembly question would require segment-arrival and completion events. Such codelets should be compared with corresponding native records, not just added to an existing feature set. Our brokers and shared capture process provide the experimental basis for that comparison; a new observation point still needs validation of its meaning, completeness, and cost.

An experiment can therefore hold the impairment and traffic settings fixed, collect the existing native observations, and add only the custom summary motivated by its question. The applied-fault record and application probe remain common reference points. This avoids confusing a change in the experimental conditions with a benefit of the added measurement.

The evaluation uses one UE on a shared VM, a fixed workload, and a low-SINR uplink with debug-enabled capture. Host scheduling, protocol adaptation, and ZMQ pacing may all contribute to delay.

\section{Related Work}
\textbf{Programmable RAN measurements.} Janus introduces verified codelets that run inline in RAN functions, enabling user-defined telemetry and control with bounded execution~\cite{janus}. Microsoft's \texttt{jrtc-apps} supplies the examples adapted here~\cite{jrtcapps}. E3/dApps provides another path to OCUDU measurement and control: its sidecar handles subscriptions, indications, and service models outside core RAN processing~\cite{dappsocudu}. These mechanisms make instrumentation programmable; our workflow supplies controlled radio conditions for comparing the observations they produce.

\textbf{Monitoring and data-driven experimentation.} SpotLight combines fine-grained RAN and platform measurements with generative models for anomaly detection and explanation~\cite{spotlight}. The Open-RAN anomaly testbed studies detection and mitigation through xApps~\cite{anomalytestbed}, while OpenRAN Gym provides data-collection, training, and testing workflows for intelligent RAN control~\cite{openrangym}. Our study is complementary: it tests which measurements describe a known intervention and its service consequence. The classifier is one use of this experimental capability, not a competing detector or a comparison of accuracies across different datasets.

\textbf{Channel emulation and fault control.} Colosseum combines software radio stacks with large-scale hardware-in-the-loop channel emulation~\cite{colosseum}. ACHEM provides a software I/Q-level channel and radio environment for USRP-based applications~\cite{achem}; OpenAirInterface includes an RF simulator with channel-model support~\cite{oairfsim}. Official srsRAN examples already insert GNU Radio processing into a ZMQ sample path~\cite{zmq4g,srsue}. Building on these established emulation approaches, we combine two broker implementations with finite directional schedules, applied/restored records, and simultaneous native, codelet, and service observations. The distinction is the integrated fault-and-measurement workflow: changing the tested impairment or observation does not require rebuilding experiment control. This focused capability complements broader emulation platforms rather than replacing their propagation models or scale.

\section{Conclusion}
We present C and Python channel brokers and a radio-fault injection workflow for direct-ZMQ 5G testbeds. The brokers change selected uplink or downlink samples; the workflow schedules and restores those changes while recording native telemetry, programmable codelets, and application outcomes. Together, they let researchers test measurement choices against known interventions without rebuilding fault control and joint capture for each question.

The interruption and stepped-noise use cases illustrate the value of this separation. Native feedback identifies the tested interruptions, whereas noise changes decoding outcomes without producing that feedback pattern. Application measurements reveal a longer slow-reply tail despite near-stable median RTTs. The selected RLC codelets describe a narrower receiving stage and leave interruption classification unchanged. These findings favor task-specific telemetry design over assuming that more features improve monitoring.

Future studies can extend the framework's evaluation to additional radio impairments, traffic loads, and multi-UE scenarios. Varying reporting intervals and collection configurations would reveal trade-offs between measurement detail, reliability, and instrumentation cost. Selected over-the-air comparisons would then assess which findings extend beyond the software testbed.

\begin{acks}
This work is partially supported by the European Union's Horizon Europe Marie Sk\l{}odowska-Curie Action SCION under Grant Agreement No. 101072375 and the NL National Growth Fund 6G flagship project Future Network Services.
\end{acks}
\balance
\bibliographystyle{ACM-Reference-Format}
\bibliography{references}
\end{document}